\documentclass[aps,prl,twocolumn,groupedaddress,showpacs]{revtex4}
\usepackage{epsfig}
\usepackage[dvipsnames,usenames]{color}
\usepackage{color}
\usepackage{amsmath}

\emergencystretch=\maxdimen
\begin{document}
\title{Universal Knight shift anomaly in the Periodic Anderson model}
\author{M.~Jiang$^{1,2}$, N.J.~Curro,$^{1}$ and R.T.~Scalettar$^1$}

\affiliation{$^1$Physics Department, University of California, Davis,
  California 95616, USA}
\affiliation{$^2$Department of Mathematics, University of California, Davis,
  California 95616, USA}

\begin{abstract}

 We report a Determinant Quantum Monte Carlo investigation which quantifies the behavior of
the susceptibility and the entropy 
in the framework of the periodic Anderson model (PAM), focussing on the
evolution with different degree of conduction electron (c) -local moment (f)
hybridization.  These results capture the behavior observed in several
experiments, including the universal behavior of the NMR Knight shift anomaly below the crossover temperature, $T^{\ast}$.
We find that $T^{\ast}$ is a measure of the onset of c-f correlations and grows with
increasing hybridization. These results suggest that the NMR Knight shift
and spin-lattice relaxation rate measurements in non-Fermi liquid
materials are strongly influenced by temperature-dependent hybridization
processes.  Our results provide a microscopic basis for the phenomenological two-fluid model of Kondo lattice behavior, and its evolution with pressure and temperature.
\end{abstract}

\pacs{71.10.Fd, 71.30.+h, 02.70.Uu}
\maketitle

Heavy-fermion materials have attracted considerable attention over the
past two decades because of their unusually large effective masses
arising from strong electron correlations
\cite{zachreview,stewartreviewRMP1984}.  These materials, which
typically contain either Ce, Yb, U or Pu ions, exhibit complex behaviors
arising from the interplay between localized and itinerant electrons.
In some cases these interactions lead to ordered ground states such as
superconductivity, antiferromagnetism, or more exotic ``hidden" order
\cite{Thompson2001,MydoshURSreview2011}.  In other cases the strong
correlations lead to a breakdown of conventional Fermi-liquid theory in
proximity to a quantum phase transition
\cite{YRSnature,ColemanQCreview,StewartHFreview}. The recent discovery
of the CeMIn$_5$ (M = Rh, Ir, or Co) class of heavy fermions, which
exhibit a broad spectrum of unusual ground states accompanied by quantum
criticality and non-Fermi liquid behavior, has highlighted the continued
need to develop a general understanding of the phase diagram of heavy
fermions, as well as a requirement to discern what behaviors are
universal rather than material-specific
\cite{tuson,Young2007,KenzelmannCeCoIn5Qphase}.

Among various experimental techniques used to investigate heavy fermion
materials, nuclear magnetic resonance (NMR) plays a central role
\cite{Curroreview}. Because the hyperfine coupling between nuclear and
electron spins  introduces an additional local effective field at the
nucleus, NMR allows one to probe the relative shift of the nuclear
resonance frequency compared with the same  nucleus in isolation. In a
normal Fermi liquid, the Knight shift is given $K = A\chi_{0}/\hbar
\gamma \mu_{B}$, where $\chi_{0}$ is the Pauli susceptibility
proportional to the density of states at the Fermi level, so that
$K\propto AN(0)$ is temperature independent. On the other hand, this
scenario fails to describe the non-Fermi liquid behavior in the normal
state of heavy fermion materials, in which the magnetic susceptibility
$\chi$ usually increases strongly with decreasing temperature. Below a
particular crossover temperature $T^{\ast}\sim 10-100$ K, the Knight
shift $K$ is no longer proportional to the magnetic susceptibility,
reflecting the onset of hybridization or lattice coherence between
conduction electron and the local moment f-electrons. This Knight shift
anomaly has been detected in all heavy fermion materials that have been
measured, including the CeMIn$_5$ family, CeCu$_2$Si$_2$, UPt$_3$, and
URu$_2$Si$_2$  ~\cite{Curro04,KentCurro}.

A variety of different hypotheses have been put forward to explain the
origin of the Knight shift anomaly, which either argue that the
hyperfine interaction acquires a temperature dependence due to Kondo
screening~\cite{theory1}, or attributes the effect to different
occupations of crystal field levels of the 4f(5f) electrons in these
materials~\cite{theory2}.  However, if the hyperfine coupling has much
larger energy scale than the Kondo and/or crystal field interactions it
is challenging to reconcile that they should give rise to the dramatic
changes observed experimentally~\cite{theory3}.

Recent progress has emerged in the context of a  two-fluid model, in
which  localized $f$-electron spins and itinerant conduction electron
spins  interact with the nuclear spins via two different hyperfine
couplings~\cite{ColemanUnderscreenedPRL1992,Pines12,twofluid1,twofluid2,twofluid3}.  The two-fluid
picture has attracted much interest as a promising phenomenological
model of several heavy-fermion behaviors, but a connection of the
predictions of this theory to a microscopic many-body Hamiltonian is
needed to provide a more comprehensive, and quantitative understanding.

It is well known that much of heavy fermion physics can be captured by
the Kondo lattice model and/or periodic Anderson model
(PAM)~\cite{SchriefferWolff} in which a lattice of f-electron local
moments is embedded into a background of conduction electrons. As the
hybridization between conduction and localized f-electrons, repulsive
interaction $U_f$ for localized moments, and the temperature are varied,
there is a competition between singlet formation by the Kondo effect and
antiferromagnetism favored by the Ruderman-Kittel-Kasuya-Yosida (RKKY)
interaction~\cite{PAMRTS}.  It is natural to consider whether these
microscopic models might also be used to understand the Knight shift
anomaly.

In this Letter, we employ the PAM to investigate the
Knight shift anomaly observed in NMR studies of several heavy fermion
materials. The half-filled two band PAM Hamiltonian reads:
\begin{eqnarray}
    {\cal H} = &-&t \sum\limits_{\langle ij \rangle,\sigma}
(c^{\dagger}_{i\sigma}c_{j\sigma}^{\vphantom{dagger}}
+c^{\dagger}_{j\sigma}c_{i\sigma}^{\vphantom{dagger}})
        -V \sum\limits_{i\sigma}
(c^{\dagger}_{i\sigma}f_{i\sigma}^{\vphantom{dagger}}+
f^{\dagger}_{i\sigma}c_{i\sigma}^{\vphantom{dagger}})
\nonumber \\
        &+& U_f \sum\limits_{i} (n^{f}_{i\uparrow}-\frac{1}{2})
(n^{f}_{i\downarrow}-\frac{1}{2})
\label{eq:PAM}
\end{eqnarray}
where $c^{\dagger}_{i\sigma}(c_{i\sigma}^{\vphantom{dagger}})$
and
$f^{\dagger}_{i\sigma}(f_{i\sigma}^{\vphantom{dagger}})$
are creation(destruction) operators for conduction and local electrons
on site $i$ and with spin $\sigma$.
$n^{c,f}_{i\sigma}$ are the associated number operators.
$t$ is the hopping amplitude between conduction electrons on the
near neighbor sites $\langle ij \rangle$ of a square lattice, $U_f$
the local repulsive interaction in the f orbital and $V$ the
hybridization between conduction and localized electrons.
We chose $t=1$ as our energy scale.
The results shown here are for a 2D square lattice but
are qualitatively unchanged in 3D, as discussed below and in
\cite{huscroft99}.

The PAM
exhibits two distinct low temperature magnetic phases \cite{PAMreference}.  For small $V$,
local $f$ moments couple antiferromagnetically via an indirect RKKY
interaction mediated in the conduction band.  At large $V$, on the
other hand, the conduction and local electrons lock into independent
singlets, and a paramagnetic spin liquid ground state forms.  This
reflects a competition between the RKKY and Kondo energy scales, $\sim
J^{2}/W$ and $\sim W e^{-W/J}$, respectively, with $J\sim V^{2}/U_f$ and
$W$ the bandwidth.

\begin{figure}
\psfig{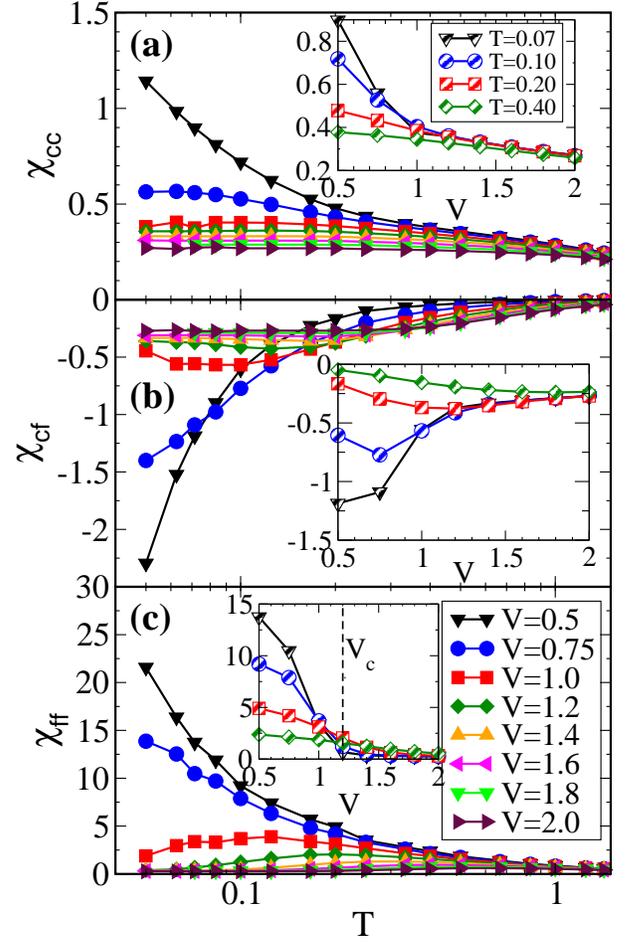} \\
\caption{(Color online) Evolution of the three components of the uniform
($q=0$) magnetic susceptibility with temperature $T$ (main panels) and
inter-orbital hybridization $V$ (insets).  At weak $V$, the conduction
and local $f$ electrons decouple, exhibiting Pauli and Curie behavior
respectively.  All three susceptibilities fall as $V$ increases and
Kondo singlets form, becoming small and temperature independent in the
vicinity of the AF-singlet transition at $(V/t)_c \sim 1.2$ (vertical
dashed line in panel (c) inset).  See [\onlinecite{PAMRTS}].  Here the
on-site repulsion of the local orbital is $U_f=4$ and the lattice size is
$12\times 12$.
\label{chis}
}
\end{figure}

We solve the PAM and address the Knight shift anomaly problem by using
Determinant Quantum Monte Carlo (DQMC) \cite{blankenbecler81}.  In this
method, a path integral expression is written for the quantum partition
function ${\cal Z}={\rm Tr\,exp}\,(\,-\beta {\cal H} \,)$, the
interaction term $n_{i\uparrow}^f n_{i\downarrow}^f$ between localized
$f$ electrons is isolated, and then mapped onto a coupling of the $f$
electron spin with a space and imaginary-time dependent auxiliary
(``Hubbard-Stratonovich") field $S_{i\tau} (n_{i\uparrow}^f
-n_{i\downarrow}^f)$.  After this replacement, which treats the
interaction energy without approximation \cite{trotter}, the fermionic
degrees of freedom can be integrated out analytically.  The result is an
exact expression for ${\cal Z}$ and operator expectation values for
spin, charge, and pairing correlation functions in terms of integrals
over the field configurations $\{ S_{i\tau} \}$.  Summing these
correlation functions over different spatial and imaginary-time
separations yields the magnetic and superfluid susceptibilities, and the
charge compressibility which signal the onset of different ordered
phases.  For the half-filled case of Eq.~\ref{eq:PAM}, the sampling is
over a positive-definite weight \cite{loh90}, and expectation values can
be obtained to low temperatures.

In the two-fluid theory~\cite{Pines12,KentCurro} the nuclear moment
$\vec I$ experiences hyperfine interactions with both the conduction and localized electron spins
$\vec S_{i}^c = (c^{\dagger}_{i\uparrow} \, c^{\dagger}_{i\downarrow})
\vec \sigma \left(\begin{array}{c} c_{i\uparrow} \\ c_{i\downarrow}
\end{array} \right) $ and $\vec S_{i}^f = (f^{\dagger}_{i\uparrow} \,
f^{\dagger}_{i\downarrow}) \vec \sigma \left(\begin{array}{c}
f_{i\uparrow} \\ f_{i\downarrow} \end{array} \right) $ via $ {\cal
H}_{\rm hyp}=  \vec{I_i}\cdot (A\vec S_{i}^c+B\vec S_{i}^f) $. Here $A$
and $B$ are the associated hyperfine couplings and include also
proportionality constants $\gamma \hbar g \mu_{B}$, and $\vec \sigma$
are the Pauli matrices.  If the electronic spins are polarized via an
external magnetic field ${\rm H}$, then
$S_{i}^c=(\chi_{cc}+\chi_{cf}){\rm H}$ and
$S_{i}^f=(\chi_{cf}+\chi_{ff}){\rm H}$, so that the magnetic
susceptibility and Knight shift are given by \begin{eqnarray}
\chi&=&\chi_{cc}+2\chi_{cf}+\chi_{ff} \nonumber \\ K &=&
A\chi_{cc}+(A+B)\chi_{cf}+B\chi_{ff}+K_{0} \label{eq:Kchi}
\end{eqnarray} respectively.   $K_{0}$ is a temperature independent term
arising from orbital and diamagnetic contributions to $K$.
If  $A\neq B$, the different
weights of the three components of the total susceptibility and their
different temperature dependencies results in
a breakdown of the linear relation between
$K$ and $\chi$ for $T < T^\ast$.

To quantify the Knight shift anomaly and the possibility of universal
behavior, we obtain the three components of the
susceptibility, $\chi_{cc}, \chi_{cf}, $ and $\chi_{ff}$, as shown in
Fig.~\ref{chis}.  When $V$ is small, the PAM describes noninteracting
conduction electrons decoupled from free moments.  At low temperature,
$\chi_{cc}$ is expected to approach a $T$-independent Pauli limit, while
$\chi_{ff}$ should have a Curie-like divergence.   This indeed
qualitatively describes \cite{footnote2} the behavior at $V=0.50$ and
$V=0.75$ in panels (a) and (c). The inter-orbital
susceptibility $\chi_{cf}$, panel (b), is negative, reflecting the
tendency of the conduction and $f$ moments to anti-align (which for
large $V$ results in singlet formation).  Note that in the singlet phase
the local, on-site contribution to $\chi_{cf}$, $\langle \, \vec S_{i}^c
\cdot \vec S_{i}^f \, \rangle$, is large.  However, because the singlets
are independent on different lattice sites, the nonlocal contributions
$\langle \, \vec S_{j}^c \cdot \vec S_{i}^f \, \rangle$ for $i \neq j$
are reduced, leading to a small $\chi_{cf}$ at large $V$.  For $U_f=4$
it is known \cite{PAMRTS} that the antiferromagnetic to singlet
transition occurs for $(V/t) \gtrsim 1.2$.  This transition is reflected
in the susceptibility components becoming temperature independent.  (See
vertical dashed line in inset to Fig.~\ref{chis}(c).)

\begin{figure}
\psfig{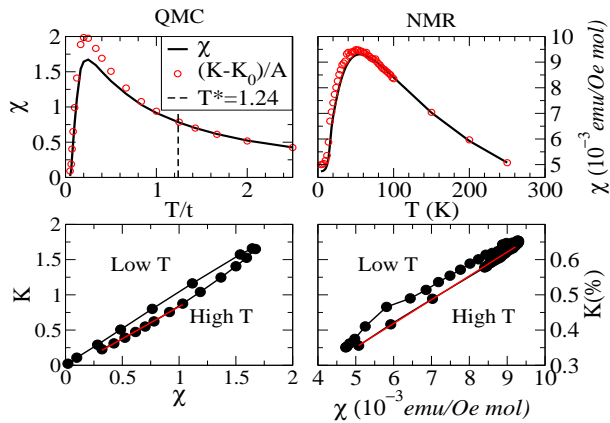} \\
\caption{(Color online) Analysis of the Knight shift anomaly.  Left
panels are DQMC data for the PAM at $V=1.2$ and $U_f=4$.  Right-hand
panels are experimental data on URu$_{2}$Si$_{2}$.
Top panels:  Susceptibility $\chi$ and renormalized Knight shift
$(K-K_0)/A$ as functions of temperature.  Above the coherence
temperature, $T>T^\ast$, $(K-K_0)/A$ tracks $\chi$.  Below $T^\ast$, the
Knight shift anomaly is evident in a deviation of $(K-K_0)/A$ from
$\chi$.  The bottom panels show $K$ versus $\chi$ with $T$ as an
implicit parameter \cite{ClogstonJaccarino}.  Both the experimental
and the
DQMC simulation data for $T>T^{\ast}$ can be fit with a straight line
$K=A\chi+K_{0}$ (red line).  The hyperfine couplings $A=0.86$, $B=2.86$
and $K_{0}=-0.056$.}
\label{kchi}
\end{figure}

Following the same procedures employed to analyze experimental data,
these DQMC results can be used to determine the coherence temperature
$T^\ast$ below which the susceptibility $\chi$ and renormalized Knight
shift $(K-K_0)/A$ break apart (top panels of Fig.~\ref{kchi}).  The
bottom panels of Fig.~\ref{kchi} show the Knight shift $K$ as a function
of susceptibility $\chi$ with $T$ as an implicit parameter. The strong
qualitative similarity between PAM simulations (left panels) and the
experimental data (right panels) in URu$_2$Si$_2$ \cite{KentCurro} is
evident. Note the $K-\chi$ plot bends counter-clockwise as $T$ is lowered, however the magnitude and direction of this effect depends on the particular magnitudes of the hyperfine couplings, $A$ and $B$. Similar plots with different values of $V$ are available in
the supplemental information.

\begin{figure}
\psfig{figure=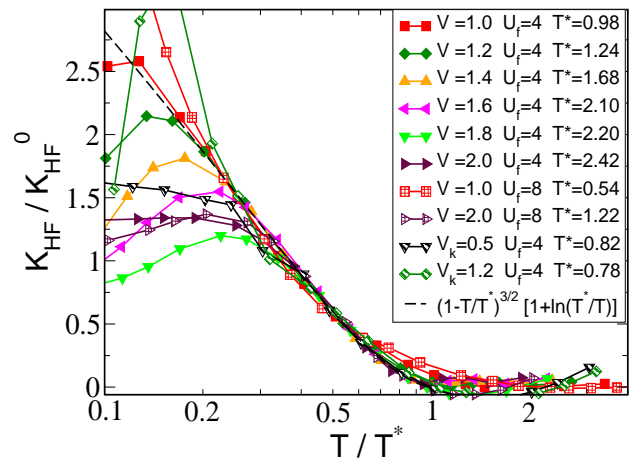,height=8.2cm,width=6.1cm,angle=-90}
\caption{(Color online) Knight shift data from DQMC simulations of the
PAM are shown to exhibit a universal logarithmic divergence with
decreasing temperature below $T^{\ast}$ in the paramagnetic state.  QMC
data are fitted for a range of high temperatures with the relation
Eq.~\ref{universal1}. Universality is seen both for different $V$ at fixed
$U_f=4$, as well as for two values of $U_f$ and
near-neighbor ($k$-dependent) hybridization $V_k (\, {\rm cos}k_x + {\rm
cos}k_y \,)$.  The breakdown of the scaling behavior
of $K_{\rm HF}$ at the lowest temperatures is also seen experimentally
and has been suggested to
arise from ``relocalization." See text for details.
We have chosen the hyperfine coupling ratio $A/B=0.3$, but the
universality is not dependent on details of the hyperfine coupling
values.  The dashed line is given by Eq. \ref{universal}.
}
\label{universal}
\end{figure}

NMR experimental results on several different families of heavy fermion compounds have revealed
that the contribution to the Knight shift from
the heavy electrons exhibits a {\it universal} logarithmic divergence with
decreasing temperature below $T^{\ast}$ in the paramagnetic
state~\cite{Curro04,KentCurro}.  The two-fluid model explains this observation
by arguing that the Knight shift component from hybridized heavy
fermions
$K_{\rm HF}=K-(A\chi+K_{0})$ is proportional to the susceptibility of the
heavy electron fluid, and can be described empirically as:
\begin{equation}\label{universal1}
  K_{\rm HF}(T)=K_{\rm HF}^{0}(1-T/T^{\ast})^{3/2}[1+\ln(T^{\ast}/T)]
\end{equation}
where $K_{\rm HF}^{0}$ and the coherence temperature $T^{\ast}$ are
material-dependent constants~\cite{YangPines}.  In Fig.~\ref{universal} we demonstrate that the predictions of this
two-fluid picture, and NMR experimental results, can also be captured in
a microscopic many-body Hamiltonian.  Specifically, if we fit our QMC
data for the Knight shift $K(T)$ in the PAM, allowing $K_{\rm HF}^0$ and
$T^\ast$ to be free parameters, we find $K_{\rm HF}(T)$ is universal over the range $0.2 T^{\ast} < T < T^{\ast}$.
Fig.~\ref{universal} shows a scaling collapse for a
range of conduction-local electron hybridizations $V$ at fixed $U_f=4$ and hyperfine couplings $A/B=0.3$.
This universality persists when $U_f$ is increased
to $U_f=8$ and for a modified form of $V$ in which
the local $f$ orbitals are hybridized with conduction orbitals on {\it
neighboring} lattice sites so that $V \rightarrow V_k \, ( \, {\rm cos}
k_x + {\rm cos} k_y \,)$.  This latter choice emphasizes the universal
scaling is insensitive to details of the band structure and bandwidth
\cite{footnote3}.  We have also verified that similar collapse behavior is
exhibited for the 3D PAM, however only smaller linear lattice sizes
$4^3$ and $6^3$ are accessible, so these data are not shown.

\begin{figure}
\includegraphics[width=\linewidth]{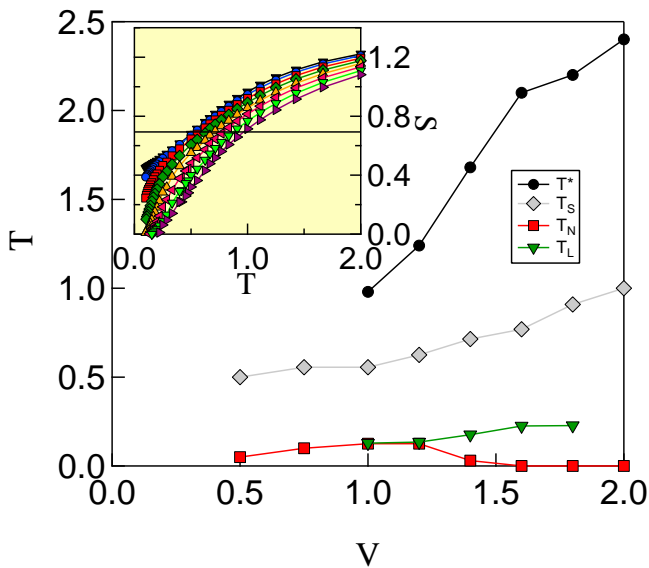}
\caption{(Color online) Evolution of $T^{*}$, $T_S$ (the temperature at
which $S=\ln 2$), $T_{loc}$ and $T_N$ with hybridization, $V$, where
$T_N$ is defined as the temperature where the antiferromagnetic
correlation length exceeds
the system size (note that $T_N=0$ in 2D).
(Inset) The thermodynamic entropy versus the temperature for different hybridization strength $V$ (symbols and colors are defined in Fig. \ref{chis}). With increasing $V$, the temperature at which the entropy decreases to the value $\ln2$ increases. This is consistent with the expectation that around the crossover temperature $T^{*}$ the hybridization between the conduction and localized f-electrons results in coherence between these degrees of freedom.}
\label{entropyT}
\end{figure}

In addition to the demonstration of universality within a microscopic model,
other features in the model agree with experimental observations. As shown in Fig. \ref{entropyT},
(i) $T^{\ast}$ increases with increasing $V$, and (ii)
the scaling behavior of $K_{\rm HF}$ breaks down
below a lower temperature $T_{\rm loc} \sim T^\ast/5$.
Although this latter behavior is not fully understood, it has been
proposed that it is associated
with the ``relocalization'' of $f$-electrons
observed in materials like CePt$_2$In$_7$
whose ground states are antiferromagnetic \cite{relocalization}.
In these materials, the finite value of $K_{\rm HF}$
at $T_{N}$ suggests that the ordered local moments remain partially
screened, emphasizing a continued competition between the heavy-fermion Kondo
liquid and a hybridized ``spin liquid'' with a lattice of local
moments associated with f-electrons~\cite{Pines12}. Because $T^{\ast}$ is an approximate measure of the onset of coherence between the itinerant and localized electrons, the two-fluid theory argues that the entropy at the crossover temperature $T^{\star}$ approaches $\ln2$ at this temperature~\cite{YangPines}. The inset of Fig.~\ref{entropyT} shows the entropy versus the temperature for different values of the hybridization $V$, and the main panel shows the evolution of $T^{*}$ and $T_S$, the temperature at which $S=\ln 2$, as a function of $V$. As expected, both temperature scales increase with increasing $V$ with the same qualitative trend.
$T_S$ is lower than $T^{\star}$ in our calculations, because $S$
includes a background contribution from free conduction electrons.
In future work we intend to develop ways 
to isolate the entropy associated with magnetic correlations,
and better test predictions\cite{Pines12} that $T_S \approx T^{\star}$.

Our DQMC simulations of the periodic Anderson model clearly capture key
features found in NMR studies of heavy fermion materials, and provide a
microscopic basis for the phenomenological two-fluid model.
Our key conclusions are that (i) the temperature evolution of the
susceptibility associated with different orbitals in this microscopic
many-body Hamiltonian results in the Knight shift anomaly as observed
experimentally; and (ii)  the Knight shift results for different choices
of interorbital hybridization and correlation energy in the localized
orbital collapse onto a universal curve.  This latter conclusion is
especially intriguing since it suggests that heavy fermion materials can
be described in a unified way, differing only through a distinct
coherence temperature, $T^\ast$, controlled by the hybridization, $V$.
Importantly, our results clearly reveal that the development of the
heavy fermion state occurs over a broad temperature range below
$T^{\ast}$, and also that both the local f-electrons as well as the
itinerant quasiparticles  contribute significantly to the NMR response
over a broad range of hybridization values where non-Fermi liquid
behavior has been observed.
Further study of the
spectral function $A(\omega)$, are in progress, and, in particular,
whether $A(\omega)$ shows any change of behavior at the coherence
temperature, as suggested recently by scanning tunneling
microscopy~\cite{STM}.

\noindent
This work was supported by
NNSA DE-NA0001842-0 and by campus-laboratory collaboration funding
from the University of California, Office of the President.
We are very grateful to David Pines, Yi-Feng Yang, and Piers Coleman
for discussions.


\bibliography{CurroBibliography}

\begin{thebibliography}{40}

\bibitem{zachreview}
Z. Fisk, D.W. Hess, C.J. Pethick, D. Pines,
J.L. Smith, J.D. Thompson and J.O. Willis,
Science 239, 33 (1988).

\bibitem{stewartreviewRMP1984}
G.R. Stewart,
Rev. Mod. Phys. 56, 755 (1984).

\bibitem{Thompson2001}
J.D. Thompson, R. Movshovich, Z. Fisk, F. Bouquet,
N.J. Curro, R.A. Fisher, P.C. Hammel, H. Hegger, M.F. Hundley,
   M. Jaime, P.G. Pagliuso, C. Petrovic, N.E. Phillips, and
J.L. Sarrao,
J. Magn. Magn. Mater. 226, 5 (2001).

\bibitem{MydoshURSreview2011}
J.A. Mydosh and P.M.  Oppeneer,
Rev. Mod. Phys. 83, 1301 (2011).

\bibitem{YRSnature}
J. Custers, P. Gegenwart, H. Wilhelm, K. Neumaier,
Y.  Tokiwa, O. Trovarelli, C. Geibel, F. Steglich, C. Pepin,
and P. Coleman,
Nature 424, 524 (2003).

\bibitem{ColemanQCreview}
P. Coleman and A.J. Schofield,
Nature 433, 226 (2005).

\bibitem{StewartHFreview}
G.R. Stewart,
Rev. Mod. Phys. 73, 797 (2001).

\bibitem{tuson}
T. Park, F. Ronning, H.Q. Yuan, M.B. Salamon,
R. Movshovich, J.L. Sarrao, and J. D. Thompson,
Nature 440, 65 (2006).

\bibitem{Young2007}
B.L. Young, R.R. Urbano, N.J. Curro,
J.D. Thompson, J.L. Sarrao, A.B. Vorontsov, and M.J. Graf,
Phys. Rev. Lett. 98, 036402 (2007).

\bibitem{KenzelmannCeCoIn5Qphase}
M. Kenzelmann, Th. Strassle, C. Niedermayer, M. Sigrist,
B. Padmanabhan, M. Zolliker, A.D. Bianchi,
R. Movshovich, E.D. Bauer, J.L. Sarrao, and J.D. Thompson,
Science 321, 1652 (2008).

\bibitem{Curroreview}
N.J. Curro, Rep. Prog. Phys. 72, 026502 (2009).

\bibitem{Curro04}
N.J. Curro, B.L. Young, J. Schmalian, and D. Pines, Phys. Rev. B70, 235117 (2004).

\bibitem{KentCurro}
K.R. Shirer, A.C. Shockley, A.P. Dioguardi, J. Crocker, C.H. Lin,
N. apRoberts-Warren, D.M. Nisson, P. Klavins, J.C. Cooley, Y.F. Yang, and
N.J. Curro,
Proc. Natl. Acad. Sci. 109, E3067 (2012).

\bibitem{YangPines}
Y.F. Yang and D. Pines, Phys. Rev. Lett. 100, 096404 (2008).

\bibitem{theory1}
E. Kim, M. Makivic, and D.L. Cox, Phys. Rev. Lett. 75, 2015 (1995).

\bibitem{theory2}
T. Ohama, H. Yasuoka, D. Mandrus, Z. Fisk, and J.L. Smith, J. Phys. Soc.
Jpn. 64, 2628 (1995).

\bibitem{theory3}
F. Mila, Phys. Rev. B40, 11382 (1989).

\bibitem{Pines12}
Y.F. Yang and D. Pines, Proc. Natl. Acad. Sci. 45, 
E3060 (2012).




\bibitem{ClogstonJaccarino}
A. M. Clogston,  and V. Jaccarino, Phys. Rev. {121},
1357 (1961).

\bibitem{ColemanUnderscreenedPRL1992}
J. Gan, P. Coleman, and N. Andrei, Phys. Rev. Lett. {68}, 3476 (1992).

\bibitem{twofluid1}
S. Nakatsuji, D. Pines, and Z. Fisk, Phys. Rev. Lett. 92, 016401 (2004).

\bibitem{twofluid2}
Y.F. Yang and D. Pines, Phys. Rev. Lett. 100, 096404 (2008).

\bibitem{twofluid3}
Y.F. Yang, Z. Fisk, H.O. Lee, J.D. Thompson, and D. Pines, Nature 454,
611 (2008).

\bibitem{SchriefferWolff}
J.R. Schrieffer and P.A. Wolff, Phys. Rev. 149, 491 (1966).

\bibitem{PAMRTS}
M. Vekic, J. W. Cannon, D. J. Scalapino, R. T. Scalettar, and R. L.
Sugar, Phys. Rev. Lett. 74, 2367 (1995).

\bibitem{huscroft99}
Carey Huscroft, A.K. McMahan, and R.T. Scalettar,
Phys. Rev. Lett. 82, 2342 (1999).

\bibitem{PAMreference} 
S. Doniach, Physica 91B, 231 (1977); B. Cornut and B. Coqblin, Phys. Rev. B5, 441 (1972). 

\bibitem{blankenbecler81}
R. Blankenbecler, D.J. Scalapino, and R.L. Sugar, Phys. Rev. D24,
2278 (1981).

\bibitem{trotter}
The `Trotter' error associated with the discretization of inverse
temperature $\beta$ is typically smaller than statistical
errors from the Monte Carlo sampling, and, in any case,
can be eliminated by extrapolation to the zero
discretization limit.

\bibitem{loh90}
 E.Y.~Loh, J.E.~Gubernatis, R.T.~Scalettar, S.R.~White,
D.J.~Scalapino, and R.L.~Sugar, Phys.~Rev.~B41, 9301 (1990).

\bibitem{footnote2}
The lack of flatness of $\chi_{cc}(T\rightarrow 0)$ expected from Pauli
behavior at the smallest hybridization, $V=0.5$, is associated
with the van-Hove singularity in the density of states at half-filling
of a tight binding, near-neighbor hopping, Hamiltonian on a square
lattice.

\bibitem{footnote3}
The choice of intra-site vs. inter-site $f$-$c$ hybridization of the PAM
fundamentally affects the noninteracting band-structure.  In the more
commonly considered intrasite case, the $U=0$ dispersion exhibits a
band-gap, while there is no gap in the inter-site case.  The effect of
these different choices has been studied in models of the `Kondo volume
collapse' in cerium, where it has been shown not to alter
the qualitative physics.  See, for example,
K.~Held, C.~Huscroft, R.T.~Scalettar, and A.K.~McMahan,
Phys.~Rev.~Lett.~85, 373 (2000).

\bibitem{relocalization}
N. apRoberts-Warren, A. P. Dioguardi, A. C. Shockley, C. H. Lin, J.
Crocker, P. Klavins, D. Pines, Y.F. Yang, and N. J. Curro,
Phys. Rev. B83, 060408 (2011).

\bibitem{STM}
P. Aynajian, E.H. da Silva Neto, A. Gyenis, R.E. Baumbach,
J.D. Thompson, Z. Fisk, E.D. Bauer, and A. Yazdani,
Nature 486, 201 (2012).

\end{thebibliography}

\end{document}